\documentclass[journal=apchd5,manuscript=article]{achemso}
\usepackage{graphicx} 
\graphicspath{Images/}
\usepackage{booktabs,multirow} 
\usepackage{xr}
\usepackage{soul}
\usepackage{comment}
\usepackage{subcaption}
\usepackage{amsmath,amsfonts,amssymb}
\usepackage{setspace}
\usepackage{tocloft}
\usepackage{arydshln}
\usepackage{lineno}
\usepackage{hyperref}
\usepackage{import}
\usepackage{caption} 
\usepackage{float} 

\title{Closing the Loop in Epitaxy with Machine Learning: Joint Optimization of Growth and Geometry in On-Chip Lasers}
\author{Mihir R. Athavale}
\email{mihirrajendra.athavale@manchester.ac.uk}
\affiliation{Department of Physics and Astronomy and The Photon Science Institute, University of Manchester, Oxford Road, Manchester, M13 9PL, United Kingdom}
\altaffiliation{Institute of Materials Research and Engineering (IMRE), Agency for Science, Technology and Research (A$^{*}$STAR), 2 Fusionopolis Way, Innovis \#08-03, Singapore, 138634, Republic of Singapore}

\author{Stephen A. Church}
\affiliation{Department of Physics and Astronomy and The Photon Science Institute, University of Manchester, Oxford Road, Manchester, M13 9PL, United Kingdom}
\altaffiliation{Centre for Engineering the Future, School of Science, Engineering and Environment, University of Salford, Salford, M5 4QJ, United Kingdom}

\author{Wei Wen Wong}
\affiliation{ARC Centre of Excellence for Transformative Meta-Optical Systems, Department of Electronic Materials Engineering, Australian National University, Mills Road, Canberra, ACT 2601, Australia}

\author{Andre KY Low}
\affiliation{School of Materials Science and Engineering, Nanyang Technological University, Singapore, 639798, Singapore}
\altaffiliation{Institute of Materials Research and Engineering (IMRE), Agency for Science, Technology and Research (A$^{*}$STAR), 2 Fusionopolis Way, Innovis \#08-03, Singapore, 138634, Republic of Singapore}

\author{Hark Hoe Tan}
\affiliation{ARC Centre of Excellence for Transformative Meta-Optical Systems, Department of Electronic Materials Engineering, Australian National University, Mills Road, Canberra, ACT 2601, Australia}

\author{Kedar Hippalgaonkar}
\affiliation{School of Materials Science and Engineering, Nanyang Technological University, Singapore, 639798, Singapore}

\author{Patrick Parkinson}
\email{patrick.parkinson@manchester.ac.uk}
\affiliation{Department of Physics and Astronomy and The Photon Science Institute, University of Manchester, Oxford Road, Manchester, M13 9PL, United Kingdom}

\begin{document}


\begin{abstract}
Achieving device-to-device reproducibility is a critical bottleneck for scalable photonic integrated circuits, as subtle variations in bottom-up epitaxial growth and fabrication severely limit yield. This study presents a machine learning workflow for III-V multi-quantum well microring lasers that first optimizes growth and geometry parameters via multi-objective Bayesian optimization, then leverages variational autoencoders (VAEs) to attribute residual device-to-device variability to its underlying sources. By explicitly targeting threshold variance alongside absolute performance, the workflow demonstrates 100\,\% lasing yield across all designs. The optimized multi-quantum well microring laser fields achieve a median lasing threshold of $16~\mu\mathrm{J}\,\mathrm{cm}^{-2}\,\mathrm{pulse}^{-1}$, a $73\%$ reduction in threshold variance relative to the previously reported best values, and a median emission wavelength of $1333~\mathrm{nm}$, in the telecommunications O-band. Furthermore, to diagnose residual performance dispersion under nominally identical conditions, VAEs are used to isolate the key components of device morphology that impact performance. This analysis successfully decouples geometric from material disorder, quantitatively linking previously unmeasured morphological variations to population-level threshold fluctuations. This data-driven workflow bridges the gap between fundamental epitaxy and reliable manufacturing, establishing a generalizable blueprint for designing and yield-optimizing complex, non-linear optoelectronic devices.
\end{abstract}


\section*{Introduction}
\label{sect:intro}  

Achieving functional uniformity across large populations of nominally identical devices is a persistent challenge in nanoscale device engineering, impacting fields spanning neuromorphic hardware\cite{Kim2021_4Kmemristor,JinMemristorReview2025} to quantum photonics\cite{Zhang2022_ScalableQDarrays,Uppu2021NatNano}. In these cases, population-level statistics rather than single-device figures of merit ultimately govern system-level performance. Integrated photonic systems exemplify this constraint: large populations of on-chip lasers whose outputs must remain within strict power and wavelength specifications are required to support dense wavelength-division multiplexing (DWDM), parallel optical interconnects, and scalable multi-laser photonic integrated circuit (PIC) architectures~\cite{Si_photonic_WDM,Roadmap_Silicon_Photonics,Miller2009Interconnects}.  For microcavity lasers and resonators, the stochastic nature of bottom-up epitaxial growth and fabrication tolerances introduce unavoidable nanometer-scale variations in geometry and material composition across a device population~\cite{Shimizu2004_MOVPEUniformity}. Consequently, process-induced mismatch in optical resonance increases the tuning, calibration, and trimming burden in complex circuits~\cite{Zortman_Si_manu,Chen2013ProcessVariation,Jayatilleka2021Trimming,thermal_tune_III_V_microdisk}.

Several approaches have been pursued to reduce performance variability in on-chip lasers. High-precision top-down fabrication can tighten wavelength distributions across nominally identical cavities \cite{Uniform_lasing_wl_microdisk}. At the circuit level, residual scatter can be mitigated using variation-tolerant resonator designs and active stabilization, including photoconductive resonator heaters \cite{Sacher2014AdiabaticMicrorings,Jayatilleka_PCH}. Bottom-up growth engineering offers a complementary route by avoiding etch-induced sidewall roughness and leveraging epitaxial selectivity, including selective area epitaxy (SAE) and aspect ratio trapping concepts that have enabled wafer-scale III--V nanoridge laser diodes on 300\,mm silicon wafers\cite{2025_De_Koninck_nanoridge_lasers,Shi2017Nanoridge}. Statistical screening and population-level analysis can characterize performance distributions, link morphology to threshold and spectral variation, and support selection and deterministic microassembly workflows such as binning and transfer printing~\cite{holistic_stephen, Alanis2017, Jevtics2020,Roelkens_uTP, Church_data_driven_discovery_2024}. Despite these strategies, addressing variability as an explicit optimization objective within the design process remains under-explored.

Data-driven optimization has emerged as a means of navigating high-dimensional and experimentally expensive photonic and metamaterial design spaces~\cite{Piccinotti2020AIPhotonics,Sakurai2019BOmetamaterials, Schul2026}, offering a natural route to incorporating population-level objectives -- such as device-to-device variability -- into the optimization target. The majority of photonics optimization and inverse-design studies focus on improving figures of merit for individual devices~\cite{Molesky2018_InverseDesignReview,Piggott2017_FabConstrainedInverseDesign}; in contrast, scaling to large device populations requires explicit control of variability arising from process variation across a wafer or within local regions~\cite{Chen2013ProcessVariation,Xing2022SpatialProcessVariations, Wang2026}. We have recently introduced a multi-objective optimization framework for multi-quantum-well microrings (MQW MRs) to reduce lasing thresholds, improve yield (defined as the fraction of nominally identical devices that exhibit lasing under the applied measurement conditions), and push the emission wavelength toward the telecommunications O-band~\cite{Athavale_MR_BO}. Here, we extend this framework by incorporating threshold variability as an explicit objective, enabling the joint optimization of (i)~the median lasing threshold, (ii)~the median lasing wavelength, and (iii)~a design-level variance quantifying the spread of the threshold distribution. Wavelength variance was not included as a dispersion objective because nominally identical MQW MR devices can lase on different cavity modes, each with a distinct emission wavelength, such that the device-to-device wavelength spread reflects both genuine structural variation and differences in mode selection, rather than structural variation alone. Using this objective set, we achieved median thresholds as low as $16~\mu\mathrm{J}\,\mathrm{cm}^{-2}\,\mathrm{pulse}^{-1}$, reduced threshold variance by $73\%$, and steered median emission wavelength to $1333~\mathrm{nm}$, while maintaining $100\%$ yield across all designs.

Despite these optimizations, residual variability can persist because both spatially systematic and stochastic process fluctuations can modulate local geometry and material properties in ways not deterministically captured by growth and design parameters~\cite{Xing2022SpatialProcessVariations,Shimizu2004_MOVPEUniformity}. We adapt a variational autoencoder (VAE) approach to diagnose the origin of residual variability in MQW\,MR lasers, going beyond previous applications of this approach used to detect drift in semiconductor processing\cite{Kim2023VAEsemiconductor,Zhu2022VAEprocessmonitoring}. Since the relevant morphological differences are subtle, a VAE was trained to learn a compressed latent representation of MQW\,MR morphology from optical images, enabling comparisons in latent space that are not accessible through direct geometric measurements alone. The morphological contribution to threshold variation is isolated independently of known growth and geometry parameters, and the added predictive value of morphology was quantified. Together, these results establish an end-to-end data-driven workflow that improves consistency through variance-aware optimization and provides morphology-aware diagnostics that isolate the independent contribution of subtle structural variation to population-level threshold fluctuations, offering a generalizable framework for optimizing complex, nonlinear optoelectronic devices.

\section*{Results and Discussion}

\subsection*{Multi-objective Bayesian Optimization for Variance Reduction}

Optimization of SAE-grown structures involves a complex interplay between electron-beam lithography used to define the growth mask and metal-organic chemical vapor deposition (MOCVD) recipes; accordingly, we describe the MQW MRs studied here in terms of a 7-dimensional growth-and-geometry parameter space. The device platform consists of bottom-up SAE-grown InAsP/InP MQW MR lasers, for which the fabrication route and chip-scale device realization have been established previously \cite{Wong_bottom_up_MQW_MRs}. The high-throughput characterization, statistical analysis, and multi-objective Bayesian-optimization (MOBO) workflow used to analyze these devices build on the approach reported previously \cite{Athavale_MR_BO,holistic_stephen}. In this framework, nominally identical devices are organized into groups termed \textit{fields}, in which all MQW MRs share the same nominal growth and geometry parameters. All devices within a field were characterized using an automated, room-temperature, power-dependent micro-photoluminescence ($\mu$PL) setup, from which lasing threshold and dominant-mode wavelength were extracted. Optical microscope images of individual MQW MRs were acquired using an independent microscopy setup for subsequent morphology analysis. Field-level performance is summarized using distribution-aware statistics, including the median lasing threshold and the median dominant-mode wavelength \cite{Athavale_MR_BO}. The training set -- representing previous rounds of growth -- comprised 95 field-level data points with 6954 individual MQW MR lasers, drawn from two batches,\cite{Athavale_MR_BO} which we refer to here as Batch~1 and Batch~2.

Field medians provide a compact measure of typical device behavior but do not capture ring-to-ring inhomogeneity within a field. To explicitly incorporate uniformity into the optimization, we included the within-field variance of the lasing threshold as an additional objective. We implemented MOBO using Gaussian-process (GP) surrogate models and jointly optimized lasing threshold, threshold variance, and emission wavelength, targeting low-threshold MQW MR lasers with uniform threshold distributions and longer-wavelength emission within the telecommunications O-band. This objective set enabled the identification of Pareto-optimal trade-offs. Candidate growth recipes and geometry settings proposed by MOBO were then experimentally implemented as Batch~3, and were fabricated and characterized using the same automated $\mu$PL workflow.

To assess reproducibility under optimized conditions, we fabricated two nominally identical copies (Batch 3A and 3B). Each optimized copy comprised four growth parameter sets (referred to as L, M, N, and O), implemented as paired samples (L1 from Batch~3A vs L2 from Batch~3B etc.). Together, these provided 72 new field-level points (nine fields per growth; full parameters are listed in the supplementary information).

\begin{figure}
\centering
\includegraphics[width=0.85\textwidth, height=17cm]
{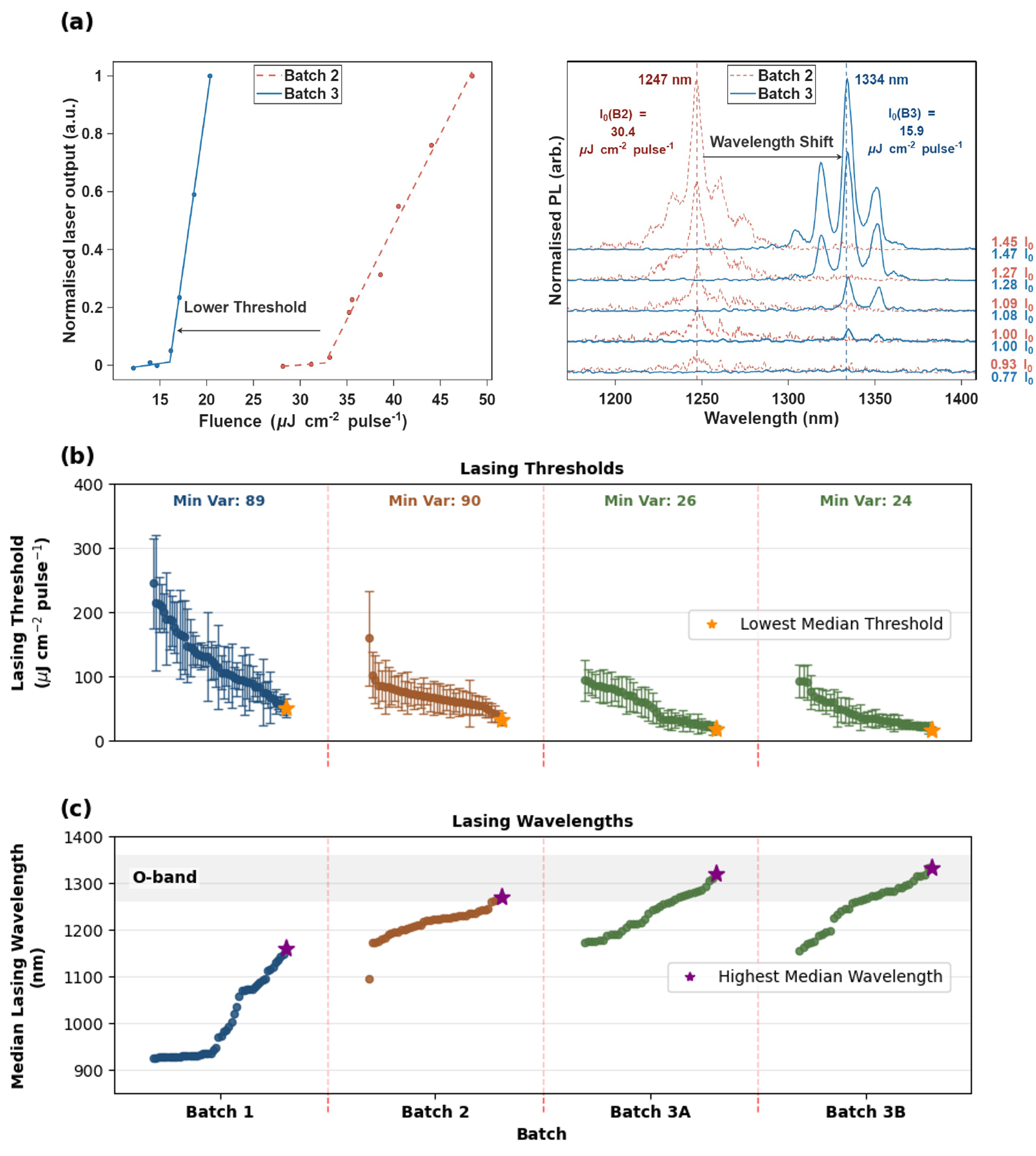}
\caption{Multi-objective Bayesian optimization improves laser performance, field-level uniformity, and emission wavelength across successive design batches. (a) Representative room-temperature $\mu$PL comparison between Batch 2 and Batch 3 devices, showing a reduced lasing threshold in Batch 3 from the light-in light-out characteristics (left) and a wavelength shift to longer emission wavelengths in the stacked spectra (right). The extracted threshold fluences $I_0$ are reduced from 30.4 to 15.9 $\mu$J cm$^{-2}$ pulse$^{-1}$. (b) Median lasing threshold for each MQW MR field across Batches 1, 2, 3A, and 3B, ordered from highest to lowest within each batch. Error bars indicate the within-field standard deviation, stars mark the lowest median threshold in each batch, and the annotated minimum within-field variance shows the progressive improvement in field-level uniformity across batches. Batch 1 and Batch 2 results were previously reported in \cite{Athavale_MR_BO}. (c) Median dominant-mode lasing wavelength for the same fields, ordered from lowest to highest within each batch. Stars mark the highest median wavelength in each batch, and the shaded region highlights the O-band spectral window.}
\label{fig:MR_BO}
\end{figure}

Figure~\ref{fig:MR_BO} summarizes field-level performance across all batches along with examples of light-in light-out (LILO) characteristics and power-dependent lasing spectra. Under the variance-aware optimization, the best field achieved a median lasing threshold of $\sim 16~\mu\mathrm{J}~\mathrm{cm}^{-2}~\mathrm{pulse}^{-1}$ ($-52\%$), a threshold variance of $24~\mu\mathrm{J}^2~\mathrm{cm}^{-4}$ ($-73\%$), and a median dominant-mode wavelength of $1333~\mathrm{nm}$ ($+65~\mathrm{nm}$), with percentage change given relative to the previous best-performing fields. Explicitly including a uniformity metric in optimization improved both typical device performance and population-level consistency, while extending emission toward communication-relevant bands.

Batch 3 outperformed the earlier batches, yet systematic differences in within-field threshold distributions were still observed (Supplementary Information). Device performance in this platform is governed by two classes of drivers: macro-scale factors, captured by the nominal 7-dimensional growth-and-geometry parameter vector, and micro-scale effects arising from spatial variations in the local growth environment, including precursor transport and surface-kinetic conditions, which are well-documented sources of morphological variation in SAE.\cite{Yuan2021_Selective} The MOBO loop operates on these growth and geometry parameters, which control the reproducible, recipe-dependent component of morphology and device performance; the residual device-to-device morphology variation within nominally identical fields is not a directly specifiable input. While MOBO navigates the macro-scale space effectively, the persistence of within-field variability under nominally identical designs therefore shows that micro-scale effects contribute to residual performance differences. Extending this framework to larger or more finely resolved parameter spaces would require additional structure, since BO scalability depends not only on the nominal parameter count but also on information gain, kernel assumptions, and whether the response is governed by a smaller number of physically relevant parameter combinations \cite{Srinivas2012,Wang_2016, Moriconi2020}.  These residual micro-scale effects motivated us to undertake a morphology-aware analysis, for which we employ a variational autoencoder (VAE) to infer a compressed latent representation of ring morphology directly from optical images.

The present study represents one completed variance-focused optimization iteration: Batch 1 and Batch 2 measurements were used to train the surrogate model, and the model-selected Batch 3 conditions constitute the experimental output of that iteration. Batch 3 has been analyzed here to assess variance reduction and morphology--threshold coupling, but these data have not yet been fed back into the model to generate a subsequent experimental batch. 

\subsection*{Interpreting Geometry in Latent Space}

\begin{figure}[H]
\centering
\includegraphics[width=\textwidth, height=13 cm]{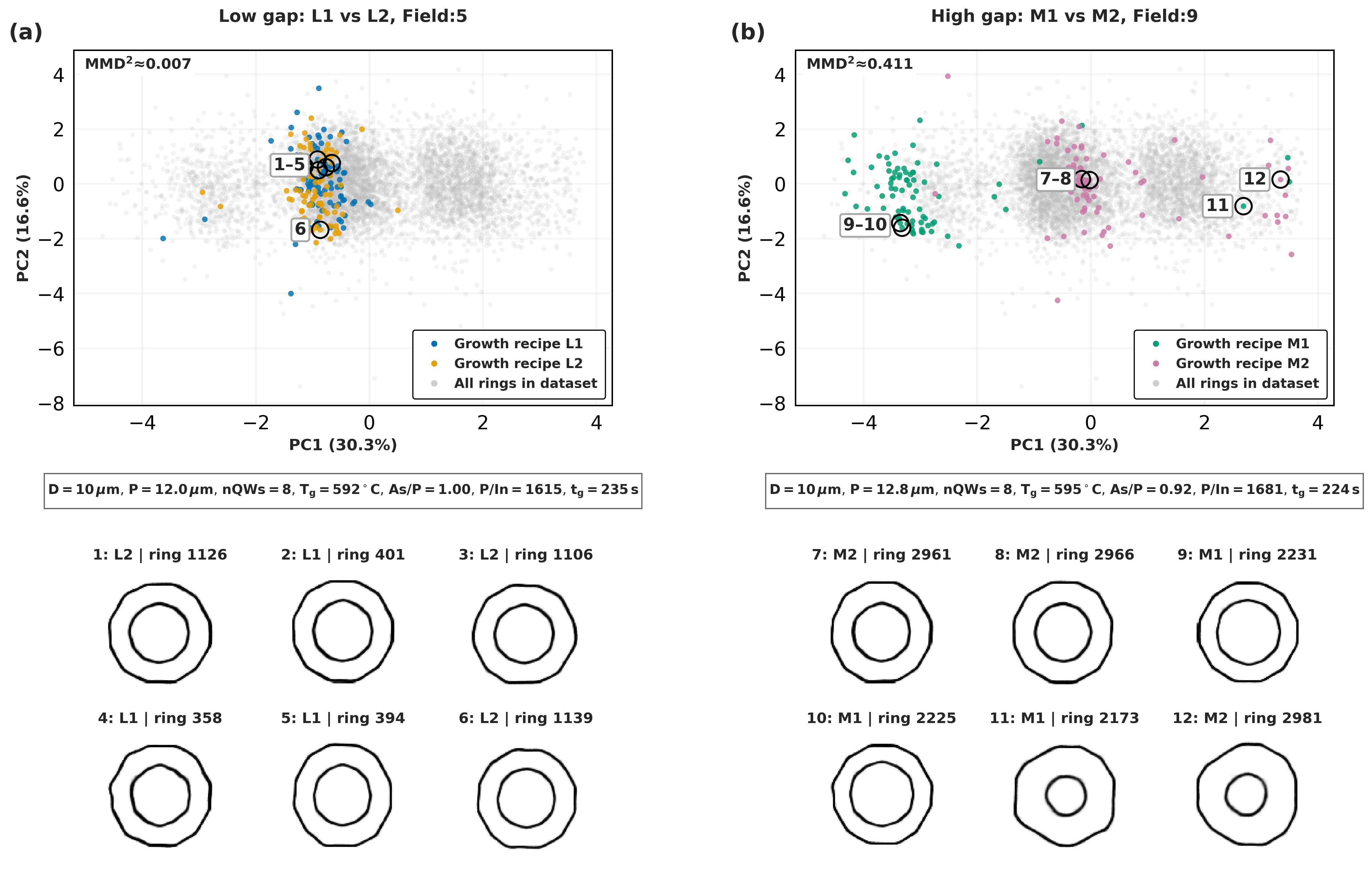}
\caption{
Latent space representations isolate and quantify subtle morphological disorder across repeat-growth subsets. (a) Two-dimensional PCA projection of the VAE latent vectors for all MQW MRs in the dataset (grey), with the globally lowest morphology-gap case highlighted (L1 vs L2, Field~5; colored). The annotated MMD$^2 \approx 0.007$ indicates that the latent distributions of the two repeat-growth subsets are nearly identical. Circled points (1--6) mark six representative rings from this field, with the corresponding reconstructed ring morphologies shown below. (b) Same global PCA axes and limits as in (a), with the globally highest morphology-gap case highlighted (M1 vs M2, Field~9; colored). The annotated MMD$^2 \approx 0.411$ indicates a pronounced distributional shift between the two subsets in latent space. Circled points (7--12) mark six representative rings from this field, with the corresponding reconstructed ring morphologies shown below. The nominal growth and geometry parameters listed below each PCA panel comprise the diameter ($D$), pitch ($P$), number of quantum wells (nQWs), quantum-well growth temperature ($T_g$), As/P ratio in the quantum well, P/In ratio during InP barrier growth, and capping growth time ($t_g$). The low-gap case exhibits closely matched morphologies across the two repeat-growth subsets, whereas the high-gap case shows more pronounced asymmetry and faceting, consistent with the larger latent space separation.
}

\label{fig:PCA_VAE}
\end{figure}

Standard geometric descriptors, such as diameter, do not capture the complex, asymmetric facet variations that evolve during SAE growth \cite{Bassett_SAE_GaAs_NW_2015}. To obtain an unbiased and compact representation of ring morphology, we trained a VAE on binarized MQW MR optical images; the VAE learns to compress each ring image into a low-dimensional latent vector that encodes its morphological features. Each MQW MR shape is represented by a 32-dimensional latent vector, allowing the morphology within a given field to be treated as a distribution of such vectors across the full MQW MR population. For each paired subset (\#1 vs \#2 for L, M, N, and O) and each field, we compared both morphology vectors and lasing threshold at the distribution level. Morphology distributions were compared in the 32-dimensional latent space using the squared maximum mean discrepancy (MMD$^2$), a metric which is suited to high-dimensional distributions via a radial basis function kernel, while threshold distributions were compared using the one-dimensional Wasserstein distance ($W_1$), appropriate for univariate data. This produces one point per field of the form (morphology gap, threshold gap), enabling a direct and unbiased test of whether larger changes in morphology are accompanied by larger shifts in threshold. Figure~\ref{fig:PCA_VAE} provides a two-dimensional Principal Component Analysis (PCA) projection of the latent space and highlights the fields exhibiting the lowest and highest morphology gaps, with representative VAE reconstructions of selected rings included to connect qualitative geometry differences to their locations in latent space.

\begin{figure}[H]
\centering
\includegraphics[width=\textwidth, height=10cm]{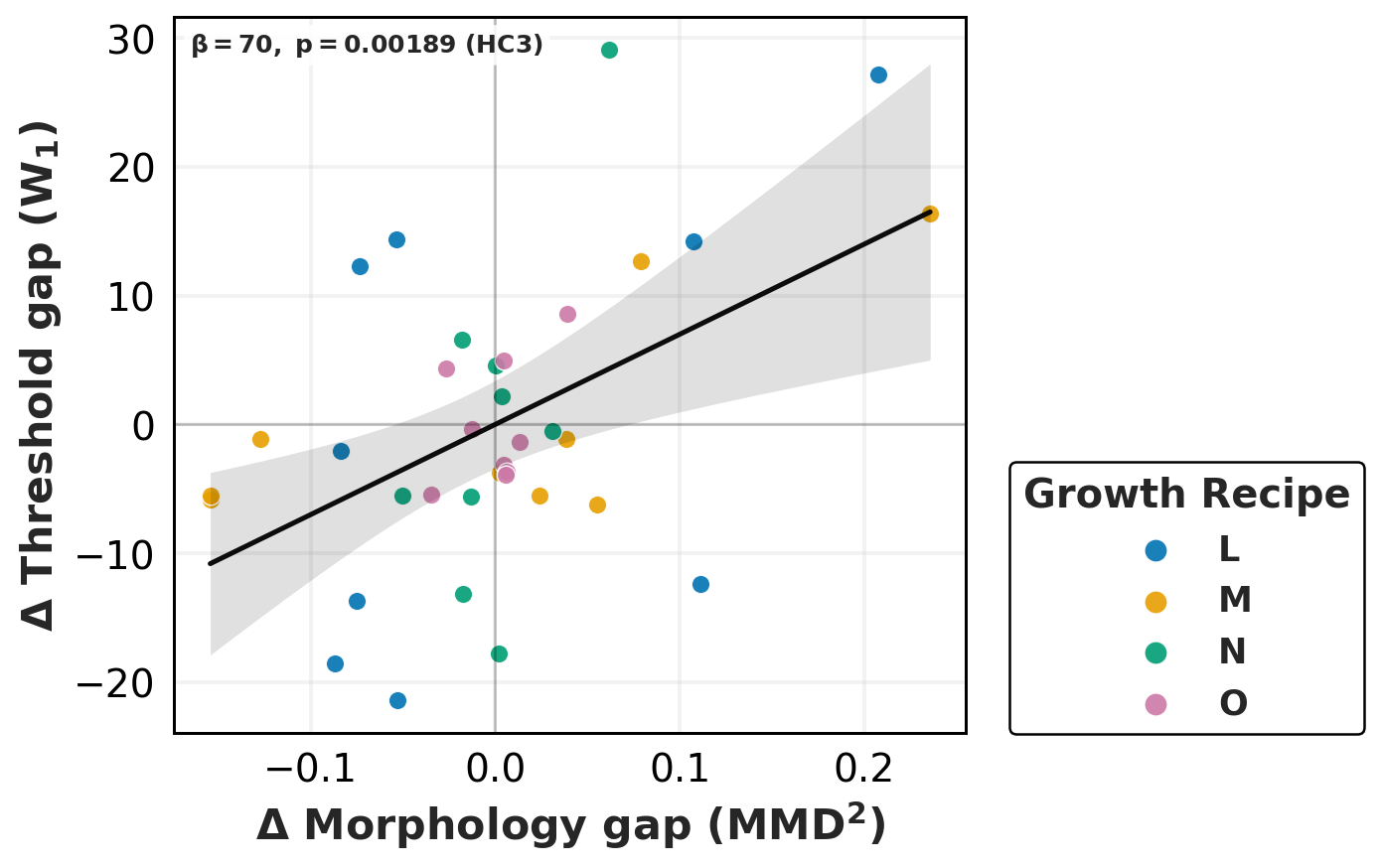}
\caption{Morphological variations directly drive performance discrepancies in nominally identical devices. Scatter plot correlating the pair-adjusted morphology gap (MMD$^2$) with the pair-adjusted threshold gap ($W_1$) for each field across paired subsets (L, M, N, and O), with both axes centered by subtracting the pair mean to remove pair-specific baselines. The black line shows the linear fit with 95\% confidence interval (shaded), and the annotated $\beta$ and $p$ value are from the centered regression using HC3 heteroskedasticity-robust standard errors.}
\label{fig:MMD2vsWSthr}
\end{figure}

For each paired subset (Batch~3A with Batch~3B counterpart), we subtracted the pair-specific mean morphology gap and pair-specific mean threshold gap, so that each point represents how much a given field differs from that pair’s average morphology gap and average threshold gap. We then fit a linear model relating the threshold gap to the morphology gap and report the slope, $\beta$, with HC3 heteroskedasticity-robust standard errors, which account for non-constant error variance across observations. As shown in Figure~\ref{fig:MMD2vsWSthr}, the pair-adjusted association between morphology gap (MMD$^2$) and threshold gap ($W_1$) was positive and statistically significant ($\beta = 70$, $p = 0.00189$, HC3), supporting the conclusion that fields with larger morphology-distribution shifts also tend to exhibit larger threshold-distribution shifts under nominally identical growth and geometry parameters. Improved control of threshold variance can be achieved with improved control of morphological variance, even within fields with 100\% yield. Robustness checks using alternative distance metrics are provided in the Supplementary Information. 

Additional threshold-relevance analysis, provided in the Supplementary Information, shows that the morphology descriptors most strongly associated with lasing-threshold variation are ring-edge radial spread, ring-edge sharpness/continuity, and facet-like asymmetry. These descriptors relate to the radial compactness, continuity, and angular uniformity of the MQW MR boundary, suggesting that optically resolvable ring-boundary deformation may influence whispering-gallery-mode confinement and increase scattering or radiative loss. This analysis should be interpreted as an association-based mechanism analysis rather than a direct causal decomposition of individual morphology features.

\subsection*{Quantifying the Determinants of Lasing Performance}

We evaluated whether VAE-derived morphology descriptors improve our prediction of ring-level lasing threshold beyond that provided by the growth recipe and nominal geometry alone. Prediction was performed using ridge regression under GroupKFold cross-validation, with folds grouped by (sample, field). Four distinct growth recipes were explored and encoded as a categorical variable (Recipe ID). We evaluated three nested models: (i) Recipe ID only, (ii) Recipe ID plus nominal size (diameter and pitch), and (iii) Recipe ID plus nominal size plus VAE latent features, enabling an additive decomposition of predictive power. To test whether any predictive gain reflects genuine morphological information rather than indirect encoding of recipe or nominal dimensions, we also evaluated transformed latent representations. In addition to the raw latents, we considered three transformed latent representations. Field-centered latents were obtained by subtracting the mean latent vector of each field from every ring in that field; in other words, we use the difference from field mean latent vector as our additional feature set for prediction. Sample$\times$field-centered latents were obtained by subtracting the mean latent vector of each $(\mathrm{sample}, \mathrm{field})$ group from every ring in that group, leaving only within-group deviations about the local mean. Parameter-residualized latents were obtained by removing the component of each latent dimension that was linearly predictable from the measured growth and geometry parameters. To avoid information leakage, field-centering and parameter residualization were performed independently within each cross-validation fold: field means were computed using only the training fold, and the parameter-to-latent Ridge model was fitted using only the training fold before being applied to the held-out data. Sample$\times$field-centering was retained as a descriptive sensitivity control rather than a strict out-of-sample predictive representation, because complete sample$\times$field groups were held out during cross-validation and their group means were therefore unavailable from the training data. Statistical significance was assessed using a permutation test in which latent vectors were shuffled within each (sample, field) group.

\begin{figure}[H]
\centering
\includegraphics[width=0.90\textwidth, height=10.5 cm]{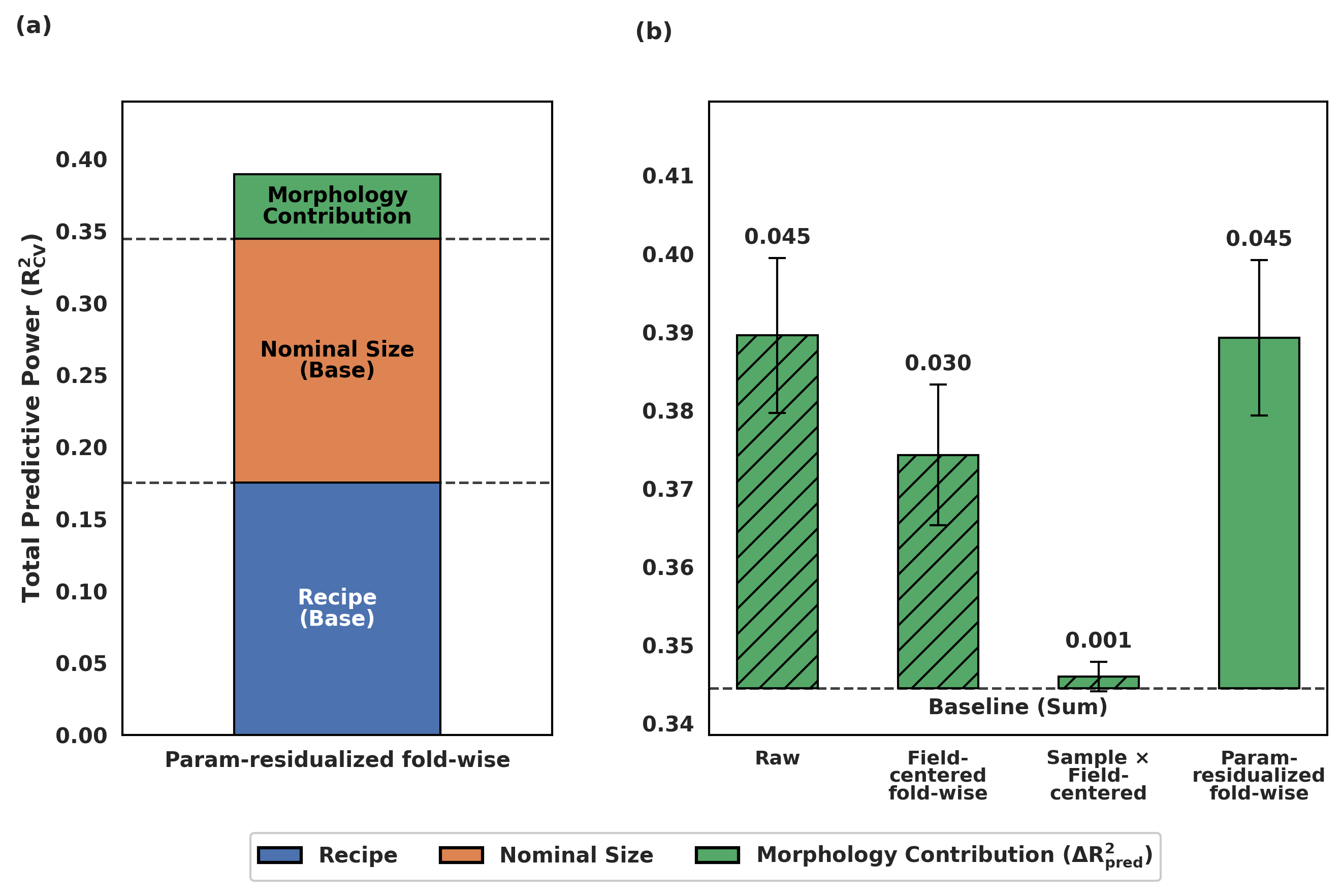}
\caption{Decomposing the drivers of lasing threshold predictability. (a) Breakdown of cross-validated predictive power ($R^2_{\mathrm{CV}}$) for the fold-wise parameter-residualized latent variant into additive contributions from growth recipe (blue), nominal ring size (orange), and VAE morphology features (green). Dashed lines mark the cumulative boundaries of the recipe-only and recipe + nominal size baselines. (b) Incremental predictive gain attributable to morphology alone ($\Delta R^2_{\mathrm{pred}}$), defined as the increase in $R^2_{\mathrm{CV}}$ when VAE latent features are added to a baseline model already containing growth recipe and nominal size. Raw latents require no fitted preprocessing. Field-centered and parameter-residualized latents were recomputed within each cross-validation fold using training-fold information only. Sample $\times$ field-centered latents are shown as a descriptive sensitivity control. Error bars denote the standard error of $\Delta R^2_{\mathrm{pred}}$ across the $N=5$ cross-validation folds, reflecting uncertainty in the estimated morphology contribution due to finite sample size. The morphology contribution remains positive for the leakage-safe raw, field-centered, and parameter-residualized representations, indicating that the added predictive value of morphology is not an artifact of globally precomputed latent preprocessing.}
\label{fig:DeltaR2_pred}
\end{figure}

Figure~\ref{fig:DeltaR2_pred} shows the resulting decomposition, with morphology-derived predictive power quantified relative to the common baseline set by recipe and nominal size. Raw latents provide a clear gain in predictive power ($\Delta R^2_{\mathrm{pred}} = 0.045$). When field-centering is implemented fold-wise, using only training-fold field means, the morphology contribution remains positive ($\Delta R^2_{\mathrm{pred}} = 0.030$). Parameter-residualized latents, computed using a parameter-to-latent model fitted only within each training fold, also retain a positive contribution ($\Delta R^2_{\mathrm{pred}} = 0.045$). These leakage-safe results show that the morphology contribution is not an artifact of globally precomputed latent preprocessing. Sample$\times$field-centering is shown as a descriptive sensitivity control, but is not treated as a strict out-of-sample predictive representation because the sample$\times$field group mean is unavailable when complete sample$\times$field groups are held out. The near-zero contribution of sample$\times$field-centered control ($\Delta R^2_{\mathrm{pred}} = 0.001$) suggests that most of the predictive morphology signal is associated with field-level or sample$\times$field-level morphology offsets rather than residual within-group device-level variation.

In contrast to the threshold study, lasing wavelength is well explained by the nominal parameter set, and morphology augmentation yields little additional predictive benefit (additional details are provided in the Supplementary Information). This may reflect the highly nonlinear and potentially non-monotonic sensitivity of lasing wavelength to subtle cavity perturbations, together with the limited ability of diffraction-limited top-view optical microscope images to resolve the fine structural variations most relevant to mode selection.

\section*{Conclusion}

This work establishes a data-driven framework that addresses a central challenge in bottom-up nanomanufacturing: improving absolute device performance while simultaneously achieving the population-level uniformity that practical integration requires. By combining MOBO with morphology-aware analysis in InP/InAsP MQW\,MR laser fields, we simultaneously reduced field-median lasing thresholds and narrowed their within-field distributions across thousands of device populations, directly targeting the reproducibility requirements of PICs, where consistent ensemble behavior is more valuable than isolated `hero' devices.

Even under nominally identical optimized growth conditions, residual threshold inhomogeneity is traceable to subtle morphological variation. A VAE trained on MQW\,MR images revealed that fields exhibiting larger morphology differences in a compressed latent space have correspondingly larger differences in lasing threshold statistics, and that latent morphology features carry significant predictive power for lasing threshold beyond measured growth and geometry parameters. Lasing wavelength, by contrast, is robustly determined by those nominal parameters, pointing to a mechanistic distinction between the two figures of merit. This distinction has a physical basis: threshold is a monotonic quantity that accumulates sensitivity to local structural imperfections, whereas lasing wavelength is governed predominantly by macroscopic resonator size and shape, with local geometry variations contributing only small perturbations. The cyclic nature of mode indexing further reduces the sensitivity of wavelength to the subtle morphological fluctuations that strongly influence lasing threshold. Together, these findings suggest that treating morphology and growth conditions as a coupled, co-optimizable design space, rather than regarding morphology as a passive outcome of growth, represents a natural extension of this framework.

Ultimately, the closed-loop optimization and quantitative morphology diagnostics established in this study represent a shift in methodology, rather than a narrow solution for a specific III-V platform or device geometry. This workflow can be adapted across nanomanufacturing contexts with performance inhomogeneity stemming from growth kinetics and structural variation, including quantum dot emitters, two-dimensional material devices, and heterogeneously integrated photonic systems; however, such cross-platform use requires platform-specific calibration data, redefined objectives and constraints, and surrogate-model retraining, rather than direct transfer of the trained model from this III--V MQW MR platform. As the density and complexity of photonic integrated circuits continue their aggressive scaling, mastering population-level reproducibility will dictate the frontier of functional design. This research provides the practical and extensible framework required to meet that challenge.


\section*{Methods}
\label{section:Methods}
\subsection*{Growth, Optical Characterization, and Imaging}

MQW MRs were grown on InP (111)A substrates by SAE using an SiO$_2$ mask defined by electron-beam lithography, followed by MOCVD growth of the InP core, the InAsP/InP MQW/barrier structure, and a final InP cap layer; TMIn, PH$_3$, and AsH$_3$ were used as precursors during growth. Lasing measurements were performed using a fluence-dependent $\mu$PL setup on a confocal microscope in reflection geometry, with a pulsed laser excitation source at 630\,nm, a repetition rate of 100\,kHz, and a pulse duration of $\sim$200\,fs. The excitation light was defocused to a diameter of 74\,$\mu$m, and the emission from a 5\,$\mu$m diameter spot was collected by an optical fiber before being analyzed using a grating spectrometer and an InGaAs array detector, with a spectral resolution of $\sim$2\,nm. Full experimental details of growth, sample fabrication, and optical characterization are available in Ref.~\cite{Wong_bottom_up_MQW_MRs}.

Optical imaging was performed separately using a Zeiss Axio Imager M2m microscope. For each field, images were acquired at 50$\times$ magnification, then cropped and matched to the corresponding entries in the dataset for subsequent morphology analysis.

\subsection*{Multi-Objective Bayesian Optimization}

MOBO was implemented in BoTorch\cite{BoTorch}, with each objective modeled by an independent SingleTaskGP surrogate trained on 95 field-level data points from prior experiments\cite{Athavale_MR_BO}. Each objective-specific surrogate was an exact GP with an RBF kernel and automatic relevance determination (ARD), so that a separate length scale was learned for each of the seven normalized input parameters. The input parameters were normalized to the predefined optimization bounds, and each output was standardized internally during model fitting. Log-normal priors were used for both the kernel length scales and the likelihood noise. The fitted hyperparameters were finite and not pinned to the imposed lower constraints, and the maximum absolute gradients after marginal likelihood optimization were small for all three objectives, indicating acceptable convergence of the GP hyperparameter optimization. The fitted length scales, likelihood noise values, final negative marginal log likelihoods, and convergence diagnostics are provided in the Supplementary Information.
qNParEGO\cite{qNEHVI_qNParEGO, ParEGO} was used as the acquisition function, converting the multi-objective problem into a single-objective one through Chebyshev scalarization with randomized weight vectors. The decision variables comprised five growth parameters (number of quantum wells, growth temperature, As/P ratio in the quantum well, V/III ratio during InP barrier growth, and capping-layer growth time) together with two geometry parameters (MQW MR diameter and the pitch–diameter difference, where pitch denotes the center-to-center spacing between neighboring MQW MRs). Four candidate recipes were selected for fabrication and characterization. To study the impact of geometry independently while holding growth parameters fixed, Latin hypercube sampling (LHS)\cite{McKay_LHS} was used to generate additional diameter and pitch–diameter difference combinations within the range spanned by the selected candidates.

\subsection*{Variational Autoencoder for Morphology Representation}
 To obtain a shape-only representation suitable for large-scale analysis, each cropped RGB MR image was converted to a binary rim-edge map using an automated pipeline in OpenCV~\cite{opencv_library}. First, the image was converted from RGB to the CIE $L*a*b$ color space and the lightness ($L$) channel (a grayscale brightness image) was used for processing. Local contrast was normalized using contrast-limited adaptive histogram equalization (CLAHE) \cite{CLAHE_1994}, and the dark MQW MR rims were emphasized using a morphological black-hat filter. The rim edges were then extracted by thresholding the black-hat response and applying light morphological cleanup to remove isolated speckles and connect small breaks. Each edge map was finally centered and uniformly scaled with a similarity transform so that the outer rim occupied a fixed fraction of a $128\times128$ frame, ensuring consistent alignment while preserving the measured morphology. 

A convolutional VAE \cite{Kingma2014VAE} was trained in PyTorch \cite{PyTorch} on $128\times128$ binarized MQW MR images. The model comprised a four-layer convolutional encoder and a symmetric transposed-convolution decoder with rectified linear unit (ReLU) activations, mapping each image to a low-dimensional latent representation parameterized by a mean vector $\mu$ and log-variance $\log \sigma^2$. Training minimized a reconstruction term given by binary cross-entropy (BCE) plus a Kullback--Leibler (KL) divergence regularization term using Adam (learning rate $10^{-3}$) for up to 200 epochs with early stopping. D4 augmentation (random $90^\circ$ rotations and flips) was applied during training. The morphology embedding for each device was taken as the latent mean $\mu$, extracted using D4 test-time augmentation and averaging. The latent dimensionality was set to $d_z = 32$ based on an ablation study maximizing downstream predictive gain ($\Delta R^2$); full selection diagnostics, together with the final model architecture, are provided in the Supplementary Information.

The resulting VAE descriptors should be interpreted as shape-normalized, optically resolvable morphology descriptors rather than direct measurements of nanoscale SAE disorder. The $128\times128$ input size defines a standardized coordinate frame for comparing ring-outline morphology, not the physical resolution of the microscope image. Features such as atomic-scale sidewall roughness, nanometer-scale SAE defects, and quantum-well thickness fluctuations are below the optical diffraction limit and are therefore not directly resolved by the optical VAE.

\subsection*{Statistical Analyses and Predictive Modeling}

 Morphological separation between matched fields in replicate MQW MR sample pairs was quantified from the 32-dimensional VAE latent vectors using two complementary metrics: the squared maximum mean discrepancy (MMD$^2$) with a radial basis function (RBF) kernel and the multivariate energy distance (ED). For the MMD$^2$ computation, the RBF bandwidth was set using the median heuristic and an unbiased estimator was used to compare latent distributions between paired subsets. Lasing threshold distributions were compared using the one-dimensional Wasserstein distance ($W_1$) and the one-dimensional energy distance. These distributional distances were computed using standard implementations available in SciPy~\cite{Virtanen2020SciPy}. Associations between morphology gaps and threshold gaps were evaluated using pair-adjusted linear regression with pair fixed effects. Heteroskedasticity-robust HC3 standard errors were used throughout~\cite{White1980,MacKinnonWhite1985}, and confidence intervals for the regression trend were obtained from the fitted model using the same robust covariance.

To quantify the predictive contribution of VAE morphology features at the MR level, ridge regression models were trained under five-fold GroupKFold cross-validation using the scikit-learn machine-learning library~\cite{Pedregosa2011ScikitLearn}, with folds grouped by $(\mathrm{sample},\mathrm{field})$. Three nested models were evaluated: (i) Recipe ID only, (ii) Recipe ID plus nominal size (diameter and pitch), and (iii) Recipe ID plus nominal size augmented with the 32-dimensional latent vectors $Z$. The morphology contribution was defined as the increase in cross-validated predictive power
\[
\Delta R^2_{\mathrm{morph}} =
R^2_{\mathrm{CV}}(\mathrm{Recipe} + \mathrm{Size} + Z)
-
R^2_{\mathrm{CV}}(\mathrm{Recipe} + \mathrm{Size}).
\]
Statistical significance was assessed using a permutation test in which latent vectors were randomly shuffled within each $(\mathrm{sample},\mathrm{field})$ group to preserve the grouping structure while disrupting any morphology--performance association. Four latent preprocessing strategies were evaluated: raw, field-centered, sample$\times$field-centered, and parameter-residualized, as defined in the results and discussion section. Raw latent vectors required no fitted preprocessing. For field-centered and parameter-residualized latents, all fitted preprocessing steps were performed independently within each cross-validation fold. Field means were computed using only the training fold, and the parameter-to-latent Ridge model was fitted using only the training fold before being applied to the held-out data. Sample$\times$field-centered latents were retained as a descriptive sensitivity control rather than a strict out-of-sample predictive representation, because complete sample$\times$field groups were held out during cross-validation and their group means were therefore unavailable from the training data.


\section*{Acknowledgements}
The Australian authors acknowledge the Australian Research Council for financial support and The Australian National Fabrication Facility, ACT Node for access to epitaxial growth and device fabrication facilities. 

\section*{Funding Sources}
UKRI Future Leaders Fellowship scheme [MR/Y03421X/1] and EPSRC (UK) grant [EP/V036343/1]. S.A.C. acknowledges Leverhulme Trust Early Career Fellowship [2025-ECF-250]. K.H. acknowledges funding from MAT-GDT Program at A*STAR via the AME Programmatic Fund by the Agency for Science, Technology and Research under Grant No. M24N4b0034 and the AISG grant number AISG3-RP-2022-028.


\section*{Conflict of Interest}
The authors declare no conflict of interest.

\section*{Data Availability}
The dataset supporting the findings of this paper is accessible at Figshare DOI: \href{https://doi.org/10.48420/27283161}{10.48420/27283161}

\section*{Code Availability}
The code used in this study is available on GitHub at: \\
\url{https://github.com/OMS-lab/Microring_Variance_Reduction}

\section*{Associated Content}
A preprint version of this work is available on arXiv at \href{https://arxiv.org/abs/2604.08390}{arxiv:2604.08390}.

\section*{Supporting Information}
Growth and geometry parameters for batch 3, Growth parameters comparison across batches, Reproducibility assessment using a reference sample, Lasing performance comparison across batches, Lasing characteristics of selected representative rings, Batch 3 reproducibility comparison, Variational autoencoder dimensionality selection, Robustness analysis of morphology--performance coupling, Predictive added value for lasing wavelength


\bibliography{references}   

\end{document}